\documentclass[aps,twocolumn,longbibliography]{revtex4-1}
\usepackage[utf8]{inputenc}
\usepackage{amsmath,amssymb,latexsym,epsfig,graphics,epsf}
\usepackage{color} 
\usepackage{dcolumn}

\usepackage[nopostdot,nogroupskip,nonumberlist]{glossaries}
\glsdisablehyper
\newacronym{LJ}{LJ}{Lennard-Jones}
\newacronym{IPL}{IPL}{inverse-power law}
\newacronym{EXP}{EXP}{exponential repulsive}
\newacronym{WCA}{WCA}{Weeks, Chandler, and Andersen}
\newacronym{AWC}{AWC}{Andersen, Weeks, and Chandler}
\newacronym{BH}{BH}{Barker and Henderson}
\newacronym{HS}{HS}{hard spheres}
\newacronym{FCC}{FCC}{face centered cubic}

\newcommand{\eq}[1]{Eq.\ (\ref{#1})}

\begin{document}
\title{Comparing four hard-sphere approximations \\for the low-temperature WCA melting line}

\author{Eman Attia}
\author{Jeppe C. Dyre}
\author{Ulf R. Pedersen}
\email{ulf@urp.dk}
\affiliation{{\it Glass and Time}, IMFUFA, Department of Science and Environment, Roskilde University, P. O. Box 260, DK-4000 Roskilde, Denmark} 
\date{\today}

\begin{abstract}
By combining interface-pinning simulations with numerical integration of the Clausius-Clapeyron equation we determine accurately the melting-line coexistence pressure and fluid/crystal densities of the Weeks-Chandler-Andersen (WCA) system covering four decades of temperature. The data are used for comparing the melting-line predictions of the  Boltzmann, Andersen-Weeks-Chandler, Barker-Henderson, and Stillinger hard-sphere approximations. The Andersen-Weeks-Chandler and the Barker-Henderson theories give the most accurate predictions, and they both work excellently in the zero-temperature limit for which analytical expressions are derived here.
\end{abstract}

\maketitle

\section{Introduction}

While systems of purely repulsive particles are rarely found in nature, they provide convenient models for both fluids and solids \cite{Hansen2013}. Examples are the \gls{IPL} systems based on a homogeneous pair potential that varies with distance $r$ as $(r/\sigma)^{-n}$ in which $\sigma$ is a length \cite{Hoover1971,Hoover1972,Heyes2007,Branka2011} and the \gls{EXP} pair potential that varies with distance as $\exp(-r/\sigma)$ \cite{EXPI,EXPII,EXPIII}. The oldest and most important purely repulsive system is that of \gls{HS} \cite{Alder1957, Wood1957, Alder1959, Alder1960}, which despite its simplicity provides a good zeroth-order model of realistic systems with both repulsive and attractive interactions \cite{Hirschfelder1954,Bernal1964,Widom1967,Barker1976,Weeks1971,Chandler1983}. A purely repulsive system has a single fluid phase and no gas-liquid phase transition. In contrast, the symmetry-breaking liquid-solid transition is present in all purely repulsive systems and, because of the absence of a gas phase, here the liquid-solid phase boundary extends to zero temperature. 

This paper studies the noted \gls{WCA} purely repulsive system  \cite{Weeks1971, Andrews1976, Chandler1983, Speedy1989, BenAmotz1990, Hess1998, BenAmotz2004, BenAmotz2004b, Heyes2006, Berthier2009, Pedersen2010, Berthier2011, Khrapak2011, Bohling2012, Dell2015, Valds2018, Chattoraj2020, Zhou2020, Attia2021, Khali2021, Banerjee2021, Toxvaerd2021, Singh2021, Heyes2021, Zhou2022, deKuijper1990, Ahmed2009, Mirzaeinia2017}, which is arrived at by cutting and shifting the \gls{LJ} interaction at its minimum \cite{Weeks1971}. In contrast to the \gls{IPL} and \gls{EXP} systems, the \gls{WCA} pair potential has a finite range beyond which pair forces are zero, like those of the \gls{HS} system. At the cutoff, the \gls{WCA} pair potential and pair forces are smooth, and at low temperatures one expects \gls{HS} approximations to apply because only insignificant ``overlaps'' are possible. Thus studies of the low-temperature melting line of the \gls{WCA} system provides an excellent testing ground for comparing different \gls{HS} approximations. This motivates the present study. In Sec. II we introduce the \gls{WCA} system and the four \gls{HS} approximations considered, as well as give a few simulation details. Section III details how we determined the \gls{WCA} melting line by interface pinning and Clausius-Clapeyron integration. The predictions of the different \gls{HS} approximations in regard to pressure and fluid/solid densities at melting are compared in Sec. IV. Finally, Sec. V provides a brief outlook. 

\section{The WCA system and hard-sphere approximations}

\subsection{The WCA system}
\begin{figure}
	\includegraphics[width=1.0\columnwidth]{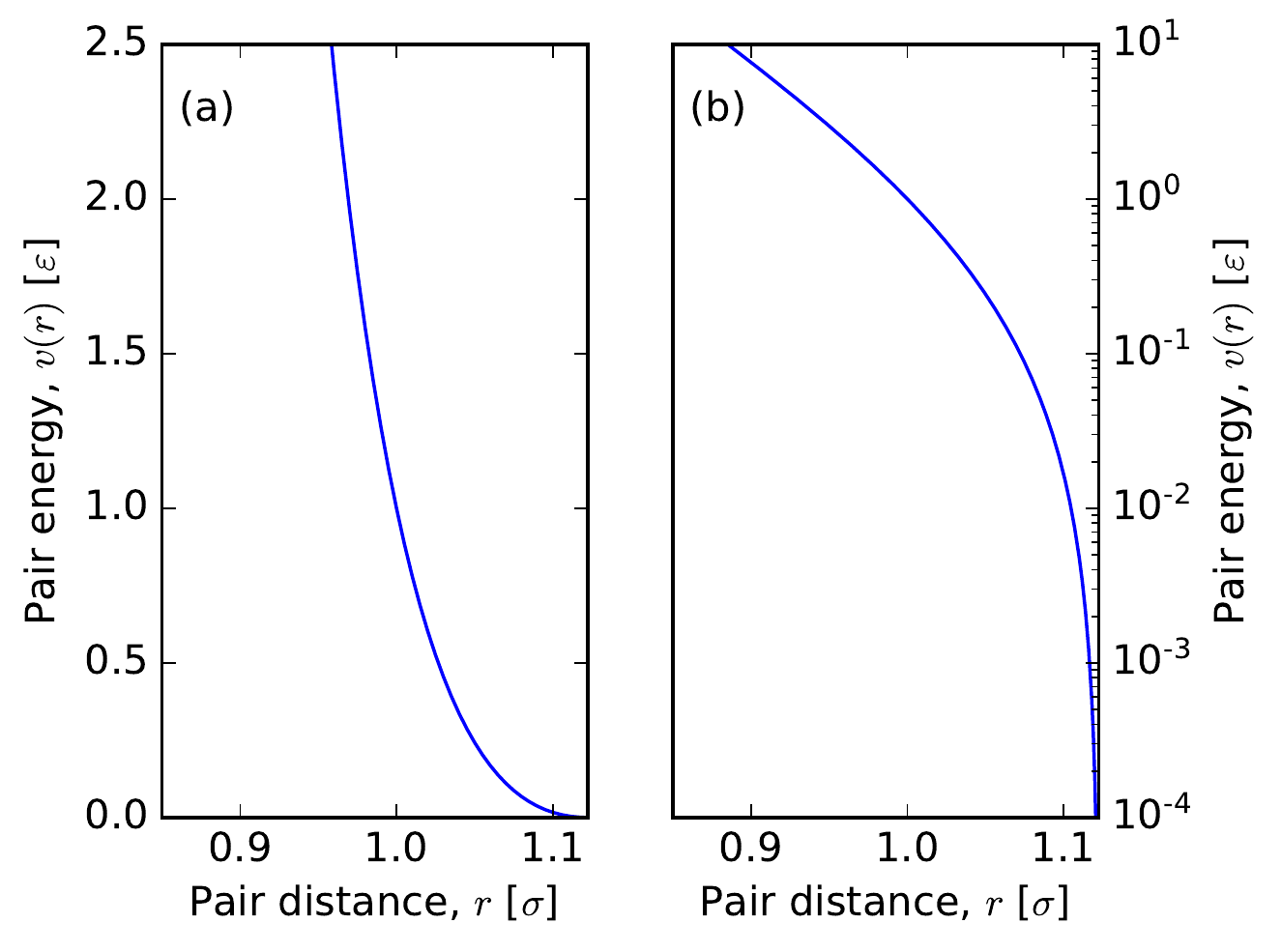}
	\caption{\label{fig:pair_potential} (a) The \gls{WCA} pair potential Eq. (\ref{eq:WCA}). (b) The same pair potential on a logarithmic energy scale, showing a steep slope at low pair energies.}
\end{figure}

We consider mono-disperse systems. Let ${\bf R}=({\bf r}_1, {\bf r}_2, \ldots, {\bf r}_N)$ be the collective coordinate vector of $N$ particles with mass $m$ confined to the volume $V$ (with periodic boundaries) and define the number density by $\rho\equiv N/V$. The potential part $\mathcal{U}({\bf R})$ of the Hamiltonian, $\mathcal{H}({\bf R})=\mathcal{U}({\bf R})+\mathcal{K}({\bf \dot R})$, is assumed to be a sum of pair-potential contributions,
\begin{equation}
	\mathcal{U}({\bf R}) = \sum_{n>m}^Nv(|{\bf r}_m-{\bf r}_n|)\,.
\end{equation}
We recall that the \gls{LJ} pair potential is defined \cite{LennardJones1924, LennardJones1924b} by
\begin{equation}\label{eq:LJ}
	v(r) \equiv 4\varepsilon \left[(r/\sigma)^{-12}- (r/\sigma)^{-6}\right]\,
\end{equation}
in which $\varepsilon$ has units of energy and $\sigma$ units of  length. The \gls{WCA} pair potential (Fig.\ \ref{fig:pair_potential}) is defined by cutting and shifting the \gls{LJ} potential at its minimum, which leads to  \cite{Weeks1971} 
\begin{equation}\label{eq:WCA}
	v(r) = 4\varepsilon\left[(r/\sigma)^{-12}-(r/\sigma)^{-6}]+1/4\right]\text{ for } r\leq r_c
\end{equation}
and zero otherwise where
\begin{equation}
	r_c = \sqrt[6]{2}\sigma\simeq1.1225\sigma
\end{equation}
The \gls{WCA} pair potential is purely repulsive since the pair force $-dv/dr\geq0$ for all $r$'s, and it is smooth since both $v(r)$ and its first derivative are continuous (the second derivative is discontinuous at $r_c$, though).
All quantities obtained by simulations are below reported in units derived from $m$, $\sigma$, $\varepsilon$ and the Boltzmann constant $k_B$.

Simulations of the \gls{WCA} system are conducted using the RUMD software package version 3.5 \cite{Bailey2017}. An initial configuration is first constructed by replicating $8\times8\times20$ \gls{FCC} unit cells, resulting in a system of $N=5120$ particles. This initial configuration is then scaled uniformly to the desired density $\rho$. If a liquid configuration is needed, the crystal is melted in a high-temperature simulation. The Newtonian equations of motion are discretized using the leap-frog algorithm \cite{Frenkel2002} with the temperature-dependent time step
\begin{equation}
	dt = 0.001\,\frac{\sigma}{\sqrt{k_BT/m}}\,.
\end{equation}
Simulations in the $NVT$ ensemble \cite{Bailey2017, Frenkel2002, GronbechJensen2019, GronbechJensen2014, GronbechJensen2014b} are realized using a Langevin thermostat with relaxation time $t_T=m/\alpha$ where $\alpha$ is a friction coefficient given by
\begin{equation}
    t_T = 0.2\,\frac{\sigma}{\sqrt{k_BT/m}}\,.
\end{equation}
For $Np_{z}T$ Langevin simulations \cite{Bailey2017, GronbechJensen2014, GronbechJensen2014b} we used the same thermostat relaxation time and the barostat relaxation time
\begin{equation}
    t_p = 100\,\frac{\sigma}{\sqrt{k_BT/m}}\,.
\end{equation}
We have found that introducing this $1/\sqrt{T}$ scaling to the relaxation times \cite{Heyes2006a} provides a simple way to ensure stability and efficiency of computations spanning several orders of magnitude in temperature (see Ref.\ \onlinecite{Ahmed2009} for a different approach). 
Note that in this way the average number of steps needed to travel the distance $\sigma$ for a thermal particle is the same for all temperatures.

\subsection{Hard-Sphere approximations to the WCA system}
Perturbation theories have proven successful for describing many fluids near freezing \cite{Zwanzig1954, Rowlinson1964,Frisch1966,Barker1967, Widom1967,Weeks1971, Andersen1971, Verlet1972, Barker1976, Andrews1976, Chandler1983, Lado1984, Kahl1989, Speedy1989, BenAmotz1990, Chang1994, Hess1998, Hansen2013, BenAmotz2004, BenAmotz2004b, Trokhymchuk2005, Heyes2006, Henderson2009, Dyre2016, Solana2019, Akhouri2022, vanWesten2022}. The basic assumption is that the pair interaction can be written as
\begin{equation}
	v(r)=v_0(r) + v_1(r)
\end{equation}
in which $v_0(r)$ is the pair potential of a (well-known) reference system and $v_1(r)$ is a small perturbation potential. Often, the \gls{HS} system is used as the reference. Several suggestions have been made for choosing the appropriate \gls{HS} diameter, $d$.
Below we list four well-known \gls{HS} criteria; in Sec. IV these are evaluated with respect to their ability to locate the solid-liquid coexistence line.

In the zero-temperature limit ($T\rightarrow0$) the \gls{WCA} pair potential approaches that of a \gls{HS} \cite{Alder1957, Wood1957, Alder1959, Alder1960} system with diameter $d=r_c$, i.e., the system described by 
\begin{equation}
	v_d(r) = \infty \text{ for } r<d
\end{equation}
and zero otherwise. While this may not be intuitively obvious since the \gls{WCA} pair potential goes smoothly to zero at the cutoff, it becomes clear when the \gls{WCA} potential is shown in a log-plot (Fig. \ref{fig:pair_potential}(b)). The simplest way of assigning an effective \gls{HS} diameter to a \gls{WCA} particle is to use the truncation distance 
\begin{equation}\label{eq:HSzero}
    d=r_c\,.
\end{equation}
This criterion is exact for $T\rightarrow0$. At finite temperatures, however, the effective \gls{HS} diameter will be smaller, and here one needs to make some physical assumption in order to improve Eq. (\ref{eq:HSzero}) to arrive at better approximations. We list below four such approximations.

\subsubsection{Boltzmann's hard-sphere criterion}
In his 1890 \emph{Lectures on Gas Theory} \cite{Boltzmann1890}  Boltzmann suggested that the effective \gls{HS} diameter $d$ should be identified with the distance of closest approach when the velocities of two head-on colliding particles are given by their average kinetic energy at far distances. This criterion can be written as
\begin{equation}
	v(d) = k_BT\,,
\end{equation}
which for the \gls{WCA} system results in 
\begin{equation}\label{eq:d_boltzmann}
	d = \frac{r_c}{\sqrt[6]{1+\sqrt{k_BT/\varepsilon}}}\,.
\end{equation}
Boltzmann's idea, which provides the simplest \gls{HS} approximation, has been used to estimate the effective \gls{HS} diameter of the \gls{WCA} system by a number of authors \cite{Andrews1976, Speedy1989, BenAmotz1990, Hess1998, BenAmotz2004b, Heyes2006}.

\subsubsection{The Andersen-Weeks-Chandler hard-sphere criterion}
A more sophisticated \gls{HS} criterion was suggested in 1971 by \gls{AWC} \cite{Andersen1971}. Their motivation was to match as well as possible the Helmholtz free energy of the pair potential in question to the associated \gls{HS} system. The \gls{AWC} criterion may be summarized as follows. If
\begin{equation}
	e(r)=\exp(-v(r)/k_BT)
\end{equation}
is the pair-potential Boltzmann probability factor, the \gls{AWC} effective \gls{HS} diameter $d$ is identified from 
\begin{equation}\label{eq:AWC}
	\int_0^\infty r^2y_d(r)\Delta e(r)dr = 0
\end{equation}
in which $\Delta e(r)=e(r)-e_d(r)$ is the blip function and $y_d(r)$ the cavity function of the \gls{HS} fluid.
In the Percus-Yevic approximation the cavity function is given analytically \cite{Wertheim1963, Kahl1989, Chang1994, Hansen2013, Trokhymchuk2005, Smith2008, Henderson2009}, which is convenient for applications of \eq{eq:AWC}. The appearance of the blip function in \eq{eq:AWC} effectively limits the \gls{AWC} integral to values near $d$. Thus it is sufficient to consider the zeroth and first shell of $y_d(r)$ to evaluate the \gls{AWC} integral of \eq{eq:AWC} with a high accuracy. We used the following implementation of the cavity function in the determination of the \gls{HS} diameter $d$ via \eq{eq:AWC} \cite{Chang1994}. If $s\equiv r/d$,
\begin{equation}
\displaystyle y_d(s) = \begin{cases}
	c_0 - c_1s + c_3s^3 & \textrm{ for } s<1 \\
	H_1(s)/s & \textrm{ for } 1 < s < 2
\end{cases}
\end{equation}
where
\begin{eqnarray}
H_1(s) 	&=& a_1\exp A(s-1) r \nonumber \\
		&+& a_2\exp B(s-1)\cos C(s-1) \\
		&+& a_3\exp B(s-1)\sin C(s-1) \nonumber \,.
\end{eqnarray}
The parameters depend on the packing fraction $\eta$ (see Eqs. (6) and (15)-(17) in Ref.\ \onlinecite{Chang1994}). For the coexistence packing fraction $\eta=0.4909$, corresponding to the density $\rho_l=0.9375\sigma^{-3}$, we have $c_0 = 58.4514$, $c_1=67.9928$, $c_3=14.3461$, $A = 1.58498$, $B = -3.68494$, $C = 3.85160$
$a_1 = 0.56770$, $a_2 = 4.23705$ and $a_3 = -1.41141$. We evaluated the \gls{AWC} integral numerically using the Python module SciPy's \cite{SciPy} implementation of QUADPACK \cite{Piessens1983}.

\subsubsection{The Barker- Henderson hard-sphere criterion}\label{sec:bh}
The \gls{BH} theory \cite{Barker1967}, which predates the \gls{AWC} theory, can be viewed as a simplification of the \gls{AWC} theory \cite{Hansen2013}. Specifically, it is assumed that $r$-squared times the cavity-function is a constant, $r^2y_d=$const., implying that Eq.\ (\ref{eq:AWC}) can be written
\begin{equation}
    0 = \int_0^\infty[1-e(r)]-[1-e_d(r)]dr\,.
\end{equation}
Since the integral of $1-e_d(r)$ is $d$, one arrives at the following \gls{HS} criterion
\begin{equation}\label{eq:bh}
    d = \int_0^\infty[1-e(r)]dr\,.
\end{equation}
The $r^2y_d=$const.\ assumption is reasonable since the blip function limits the integral to values near $d$ where $y_d$ does not change much when the temperature is sufficiently low. As $T$ is lowered, the blip function narrows; thus the \gls{AWC} diameter reduces to the \gls{BH} criterion when $T\rightarrow0$. Note that the \gls{BH} criteria depends on temperature but not on density (the \gls{AWC} criteria depends on both temperature and density). The \gls{BH} integral of \eq{eq:bh} is easily evaluated numerically using, e.g., the Python module SciPy's \cite{SciPy} implementation of QUADPACK \cite{Piessens1983}.

\subsubsection{Stillinger's hard-sphere criterion}
At low temperatures, the integrand of the \gls{BH} criterion Eq.\ (\ref{eq:bh}) changes rapidly from nearly unity for $r<d$ to nearly zero for $r>d$. This motivates the \gls{HS} criterion proposed by Stillinger in 1976 \cite{Stillinger1976, Likos2001, Heyes2021}. He pragmatically identified the \gls{HS} diameter from where the pair-potential Boltzmann factor equals one half, i.e., 
\begin{equation}\label{eq:d_stillinger_assumption}
    e(d)=\frac{1}{2}\,.
\end{equation}
Stillinger introduced this in connection with his study of the Gaussian-core model \cite{Stillinger1976}. The same idea can also be applied to the \gls{WCA} potential, however, leading \cite{Heyes2021} to 
\begin{equation}\label{eq:d_stillinger}
    d = \frac{r_c}{\sqrt[6]{1+\sqrt{k_BT\ln(2)/\varepsilon}}}\,.
\end{equation}
The functional form of this \gls{HS} criterion is identical to that of Boltzmann if $T$ is replaced by $T\ln(2)$: The factor 2 is here from Eq.\ (\ref{eq:d_stillinger_assumption}), and with $e(d)=1/\exp(1)$ one arrives at Boltzmann's criterion \cite{BenAmotz2004b}.

\begin{figure}
	\includegraphics[width=1.0\columnwidth]{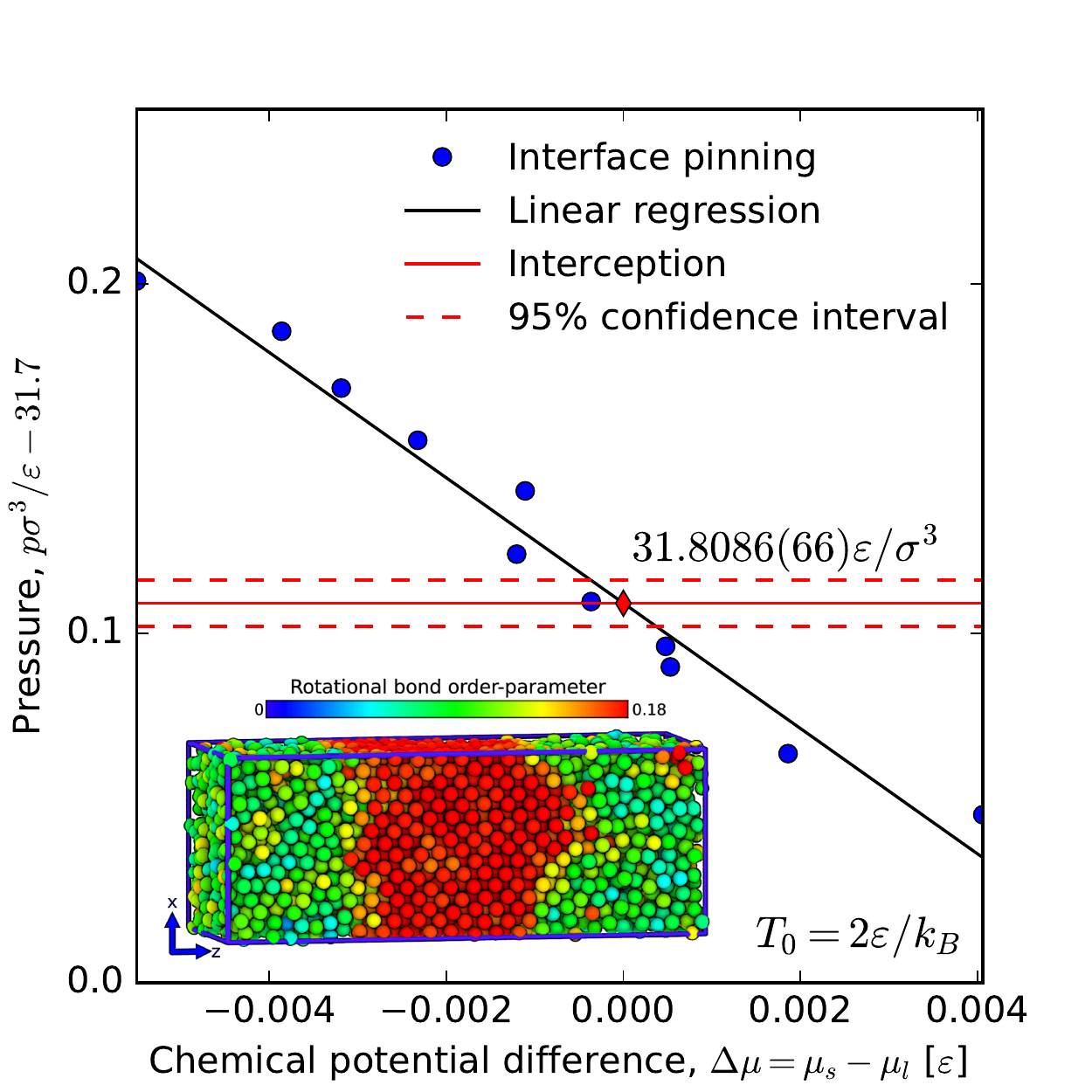}
	\caption{\label{fig:interface_pinning} Determination of the coexistence pressure at the temperature $T_0=2\varepsilon/k_B$ (red diamond) by means of the interface-pinning method \cite{Pedersen2013, Pedersen2013b, Thapar2014, Pedersen2015, Cheng2019, Newman2019, Steenbergen2017, Sharma2018, Zou2020, Zhu2021}. See the text for details. The inset shows an interface-pinned configuration where the colors indicate the rotational bond order parameter $\bar q_4$ defined in Ref.\ \onlinecite{Lechner2008}. With this coloring crystalline particles are reddish, while fluid particles are greenish.}
\end{figure}

\begin{table*}
	\caption{\label{table:phase_transition} Selected state points on the coexistence line determined with the interface pinning (IP) method and by numerical integration of the Clausius-Clapeyron (CC) relation (the Supplementary Material gives all computed data). The numbers in parenthesis give the statistical uncertainty (95\% confidence interval).}
	\begin{ruledtabular}
		\begin{tabular}{d|dddc}
			T \: [\varepsilon/k] & p \: [\varepsilon/\sigma^3] & \rho_l \: [1/\sigma^3] & \rho_s \: [1/\sigma^3] &  Method \\
			\colrule
			20 & 633.309 & 1.78328 & 1.85850 & CC \\
			2 & 31.8086(66) & 1.08441(5) & 1.15192(6) & IP \\
			2 & 31.7532 &     1.08413 &    1.15163 & CC \\
			0.2 & 2.05169(33) & 0.80004(3) & 0.87356(4) & IP \\
			0.2 & 2.05118 &     0.79992 &    0.87358    & CC \\
			0.02 & 0.174944(47) & 0.70638(5) & 0.77889(6) & IP \\
			0.002 & 0.016687(3) & 0.67717(3) & 0.74791(3) &  IP \\
			0.002 & 0.016680 &    0.67705    & 0.74792    & CC
		\end{tabular}
	\end{ruledtabular}
\end{table*}

\begin{figure}
	\includegraphics[width=1.0\columnwidth]{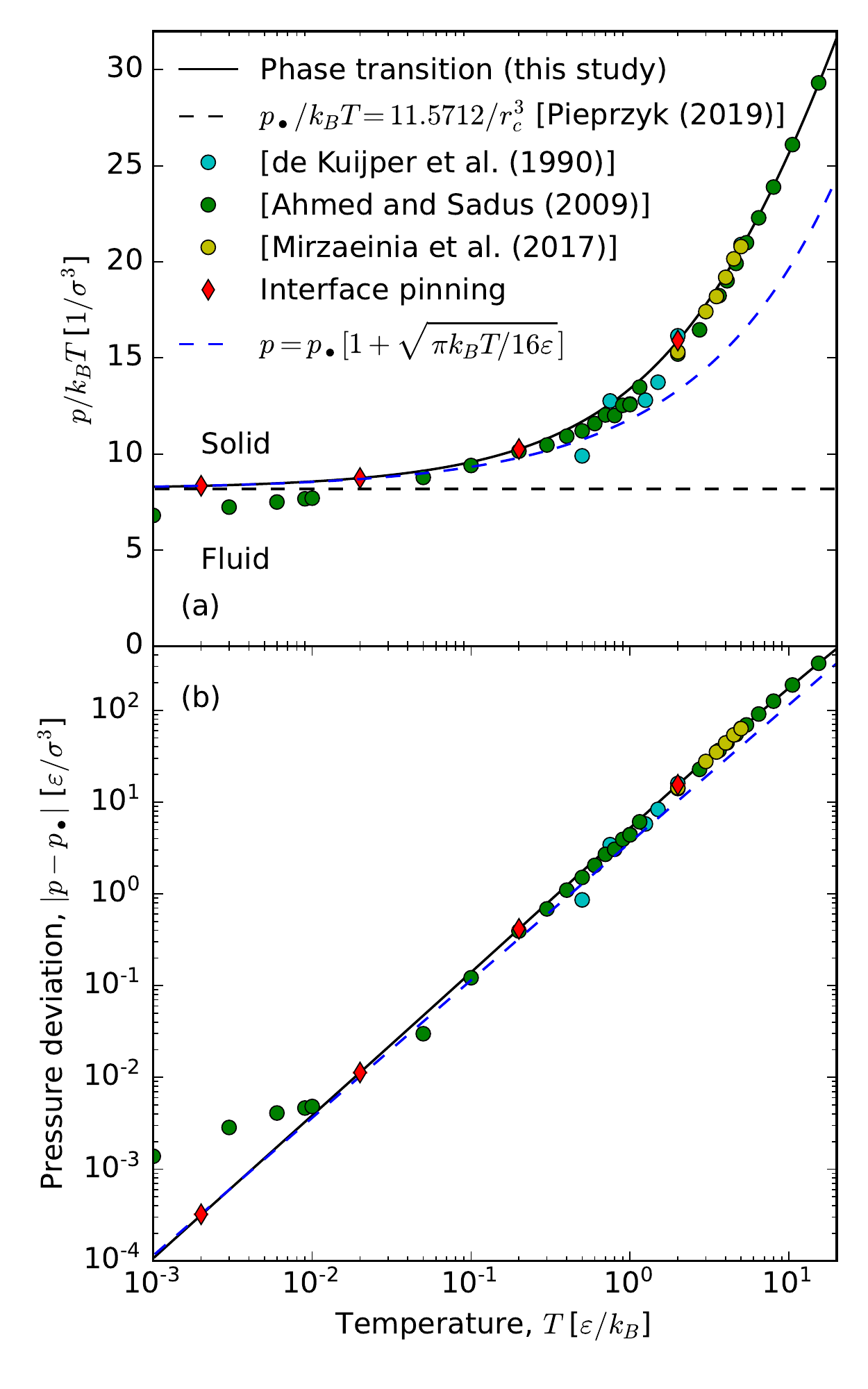}
	\caption{\label{fig:coexistence_line_compare} Coexistence pressure as a function of the temperature. (a) The solid black line shows the reduced coexistence pressure $p/k_BT$ as a function of the temperature (this study). The black dashed line gives the $T\rightarrow0$ \gls{HS} limit, $p_\bullet/k_BT$,  and the colored dots represent literature coexistence pressures  \cite{deKuijper1990, Ahmed2009, Mirzaeinia2017}. The red diamonds were computed with the interface-pinning method (this study). The blue dashed line shows that at low temperatures the pressure scales as $T^{3/2}$, as expected from \gls{HS} theories (see the text). (b) The absolute value of the coexistence pressure in excess of its $T\rightarrow 0$ limit.}
\end{figure}

\begin{figure}
	\includegraphics[width=1.0\columnwidth]{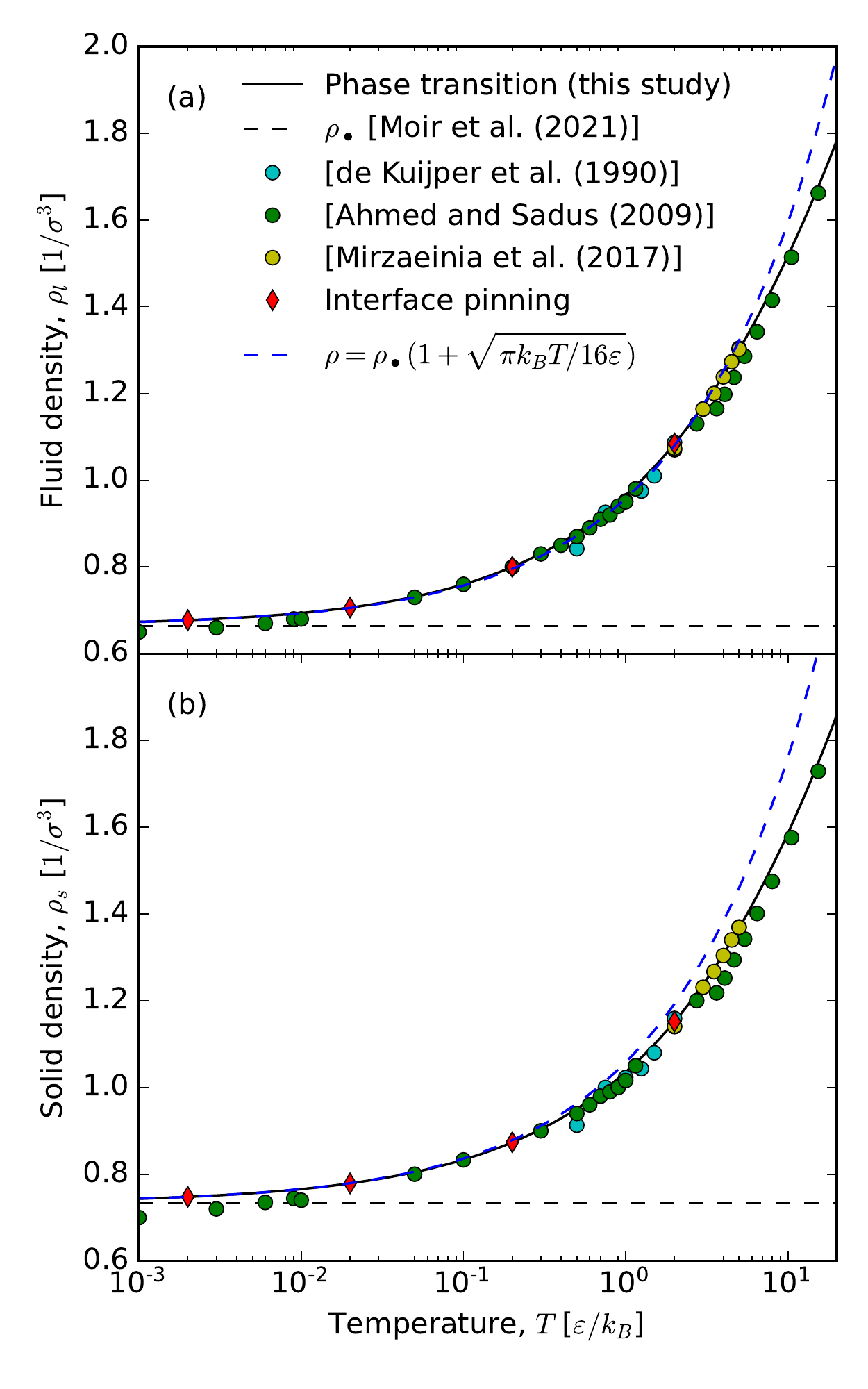}
	\caption{\label{fig:coexistence_line_compare_rhoT} Fluid density at freezing and solid density at melting as functions of the temperature. (a) The solid black line shows the density of the fluid at coexistence (this study). The dashed line marks the $T\rightarrow0$ limit and the colored dots are literature data \cite{deKuijper1990, Ahmed2009, Mirzaeinia2017}. The red diamonds are densities computed with the interface-pinning method. (b) The solid black line shows the density of the solid at coexistence (this study), the dashed line is the $T\rightarrow0$ limit, and the colored dots represent literature data \cite{deKuijper1990, Ahmed2009, Mirzaeinia2017}.  }
\end{figure}

\section{Numerical determination of the phase transition line}

The interface pinning method \cite{Pedersen2013, Pedersen2013b, Thapar2014, Pedersen2015, Cheng2019, Newman2019, Steenbergen2017, Sharma2018, Zou2020, Zhu2021} is used to compute the solid-liquid chemical potential difference $\Delta \mu$ for isothermal state-points at temperatures 0.002$\varepsilon/k_B$, 0.02$\varepsilon/k_B$, 0.2$\varepsilon/k_B$, and 2$\varepsilon/k_B$. For a given temperature, we first set up a \gls{FCC} crystal elongated in the $z$-direction with the given density and compute the equilibrium pressure in an $NVT$ simulation. We then construct a configuration of half-crystal and half-fluid by a high-temperature simulation, where particle positions are only updated for half of the particles (resulting in melting for these particles). This produces a configuration similar to the one shown in the inset of Fig.\ \ref{fig:interface_pinning}. We then perform an $Np_zT$ simulation by adding a harmonic bias-field to the potential part of the Hamiltonian,
\begin{equation}
    \mathcal{U}_\textrm{IP}({\bf R}) = \mathcal{U}({\bf R}) + \frac{\kappa}{2}\left(Q({\bf R})-a\right)^2,
\end{equation}
which forces the system toward configurations with a fluid-crystal interface. 
Here, $\kappa$ and $a$ are parameters of the bias-field, and $Q({\bf R})$ is an order-parameter that measure crystallinity though long-range order (see Eq.\ (15) in Ref.\ \onlinecite{Pedersen2013}). The chemical potential difference between the two phases, $\Delta\mu$, is computed from the average force, $\kappa(\langle Q({\bf R})\rangle - a )$, which the bias field results in on the system (see Eq.\ (9) in Ref.\ \onlinecite{Pedersen2013}). This is then repeated for several \gls{FCC} densities (and thus pressures) near coexistence. As an example, Fig.\ \ref{fig:interface_pinning} shows the pressures versus the computed chemical potentials at $2\varepsilon/k_B$, considering eleven pressures slightly above $31.7\varepsilon/\sigma^3$. The coexistence state point at $\Delta \mu=0$ is determined by linear regression, compare the solid line on Fig. \ref{fig:interface_pinning}. From this we find the coexistence pressure  $p=31.8086(66)\varepsilon/\sigma^3$ where the number in parenthesis gives the statistical error on the last two digits using a 95\% confidence interval. Table \ref{table:phase_transition} reports the thermodynamic coexistence data obtained by the interface-pinning (IP) method and numerical integration of the Clausius-Clapeyron (CC) relation as detailed below.

While the interface-pinning method is accurate and provides specific error estimates, it can be computationally expensive because long simulations are needed to properly represent interface fluctuations, which are usually significantly slower than fluctuations of the bulk solid and fluid \cite{Pedersen2013}. As an alternative, we determine most points on the coexistence line by numerical integration of the Clausius-Clapeyron relation (below $s$ and $v$ are the entropy and volume per particle)
\begin{equation}\label{eq:clausius_clapeyron}
\frac{dp}{dT} = \frac{\Delta s}{\Delta v}\,.
\end{equation}
This is an example of the Gibbs-Duhem integration methods discussed by Kofke \cite{Kofke1993, Kofke1993b}, which do not involve slow fluctuations of an interface. The volume difference $\Delta v=v_l-v_s$ and the entropy difference $\Delta s=s_l-s_s=(\Delta u + p\Delta v - \Delta \mu)/T$ can both be evaluated from standard $NpT$ simulations of the two bulk phases at coexistence (where of course $\Delta\mu=0$).

We use a trapezoidal predictor-corrector method to compute coexistence pressures at the temperatures $T_i=0.02\times10^{(i/24)}$ where $i$ is an integer, compare the solid black line on Fig.\ \ref{fig:coexistence_line_compare}. Substituting $t=T$ and $y=p$ we can write the first-order differential equation to be solved in the standard form
\begin{equation}
y' = f(t, y)
\end{equation}
where $f$ is the slope evaluated as $\Delta s/\Delta v$ 
(Eq.(\ref{eq:clausius_clapeyron})). Suppose one knows the point $(t_i, y_i)$ on the coexistence line, either from the interface-pinning method or from a previous step of the  Clausius-Clapeyron integration, and wish to compute the next point $(t_{i+1}, y_{i+1})$. Let the step length in $t$ be $h=t_{i+1}-t_{i}$. The prediction of the simple Euler algorithm is
\begin{equation}
y^{(0)}_{i+1} = y_{i} + hf(t_i, y_i)\,.
\end{equation}
A better estimate is provided using Heun's method: 
\begin{equation}
y^{(1)}_{i+1} = y_{i} + \frac{h}{2}[f(t_i, y_i)+f(t_i+h, y^{0}_{i+1})]
\end{equation}
The next estimate in an iterative predictor-corrector approach is
\begin{equation}
y^{(2)}_{i+1} = y_{i} + \frac{h}{2}[f(t_i, y_i)+f(t_i+h, y^{1}_{i+1})]
\end{equation}
or, in general,
\begin{equation}
y^{k+1}_{i+1} = y_{i} + \frac{h}{2}[f(t_i, y_i)+f(t_i+h, y^k_{i+1})]\,.
\end{equation}
In the limit of large $k$'s this converges to the trapezoidal rule of integration where forward and backward integrations yield the same result.

Which criterion to use in order to determine when the predictor-corrector iterations have converged? To answer this we note that since slopes are evaluated from finite $NpT$ simulations, we expect a significant statistical error on the $f$'s used above. If $\bar f(t, y)$ is the theoretical slope, $f(t, y)=\bar f(t, y)+e_f$ where $e_f$ is drawn from a normal distribution with standard deviation $\sigma_f$. This error is estimated by dividing $NpT$ simulations into statistically independent blocks \cite{Flyvbjerg1989}.
The error on $y^{k+1}_{i+1}$ is $e_y=he_f$ and $\sigma_y=|h|\sigma_f$. It is sensible to terminate the predictor-corrector iteration when
\begin{equation}
|y^{k+1}_{i+1}-y^{k}_{i+1}|<\sigma_y\,
\end{equation}
since this indicates that changes of $y_{i+1}$'s are mainly due to the statistical uncertainty on the slopes.

In summary, numerical integration of the Clausius-Clapeyron relation comes with errors from ignoring higher-order terms and from the statistical uncertainty of the slopes. To quantify the overall error of the integration we can compare to the accurate estimates from interface pinning at selected state points. As an example, for $T_{48}=2\varepsilon/k_B$ from the Clausius-Clapeyron integration we estimate the coexistence pressure to be $31.7532\varepsilon/\sigma^3$, which should be compared to $31.8086(66)\varepsilon/\sigma^3$ for the interface-pinning method, see Table \ref{table:phase_transition}. The error of the computed phase-transition line is not visible in most figures of this paper, with  notable exceptions at low temperatures (error bars are shown in the below figures whenever errors are significant).

Figures \ref{fig:coexistence_line_compare} and \ref{fig:coexistence_line_compare_rhoT} show coexistence pressures and densities, respectively, from this study and from the literature \cite{deKuijper1990, Ahmed2009, Mirzaeinia2017}. We note that the low-temperature estimates of Ref.\ \onlinecite{Ahmed2009} are not accurate, while the high-temperature estimates of Refs.\  \onlinecite{deKuijper1990, Ahmed2009, Mirzaeinia2017} are consistent with our results. As a consistency check, we note that the computed coexistence line reaches the \gls{HS} limit \cite{Pieprzyk2019} when $T\rightarrow0$ (the dashed lines on Figs.\ \ref{fig:coexistence_line_compare} and \ref{fig:coexistence_line_compare_rhoT} show the \gls{HS} limits). 

\begin{figure}[!ht]
	\includegraphics[width=1.0\columnwidth]{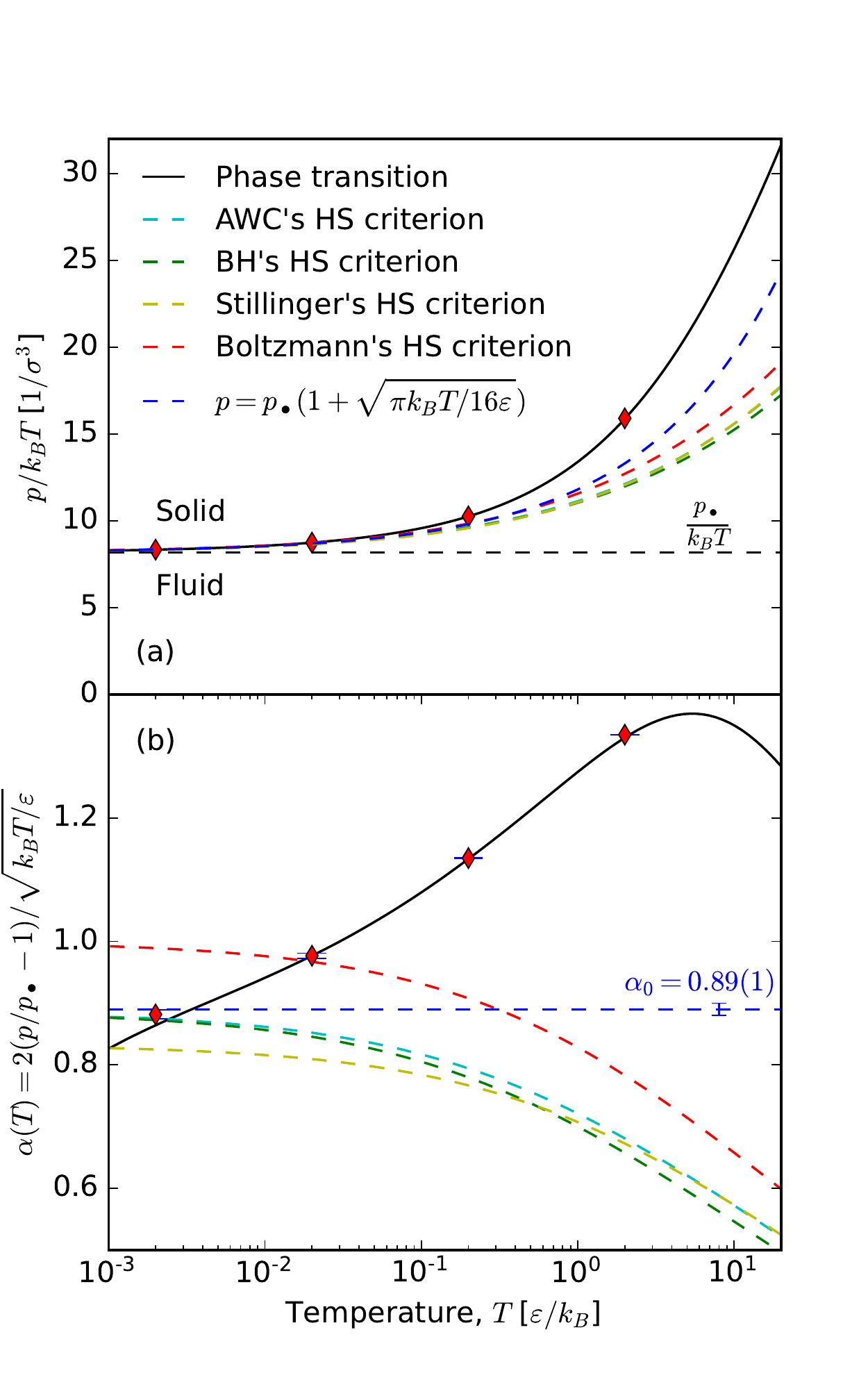}
	\caption{\label{fig:HS_theory}Melting-line pressure compared to \gls{HS} predictions.
	(a) The solid black line shows the reduced coexistence pressure, $p/k_BT$. The dashed lines show predictions of the \gls{HS} theories (see text for details). The red diamonds show coexistence pressures computed with the interface-pinning method. (b) $\alpha_p(T)=2(p/p_\bullet-1)/\sqrt{k_BT/\varepsilon}$ (Eq.\ (\ref{eq:alpha})) along the computed phase transition line (black solid) and the theoretical predictions also shown in the upper panel (dashed lines). The blue dashed line ($\alpha_0=0.89(1)$) is the $T\rightarrow0$ limit determined from coexistence densities, see Fig.\ \ref{fig:coexistence_line_rhoT}(b). \gls{AWC} and \gls{BH} give accurate predictions in the low-temperature limit. The red diamonds are the results of the interface-pinning method where blue error bars indicate the statistical error. We note a systematic inaccuracy of the Clausius-Clapeyron integration (solid black) at the lowest temperatures.}
\end{figure}

\begin{figure}[!ht]
	\includegraphics[width=1.0\columnwidth]{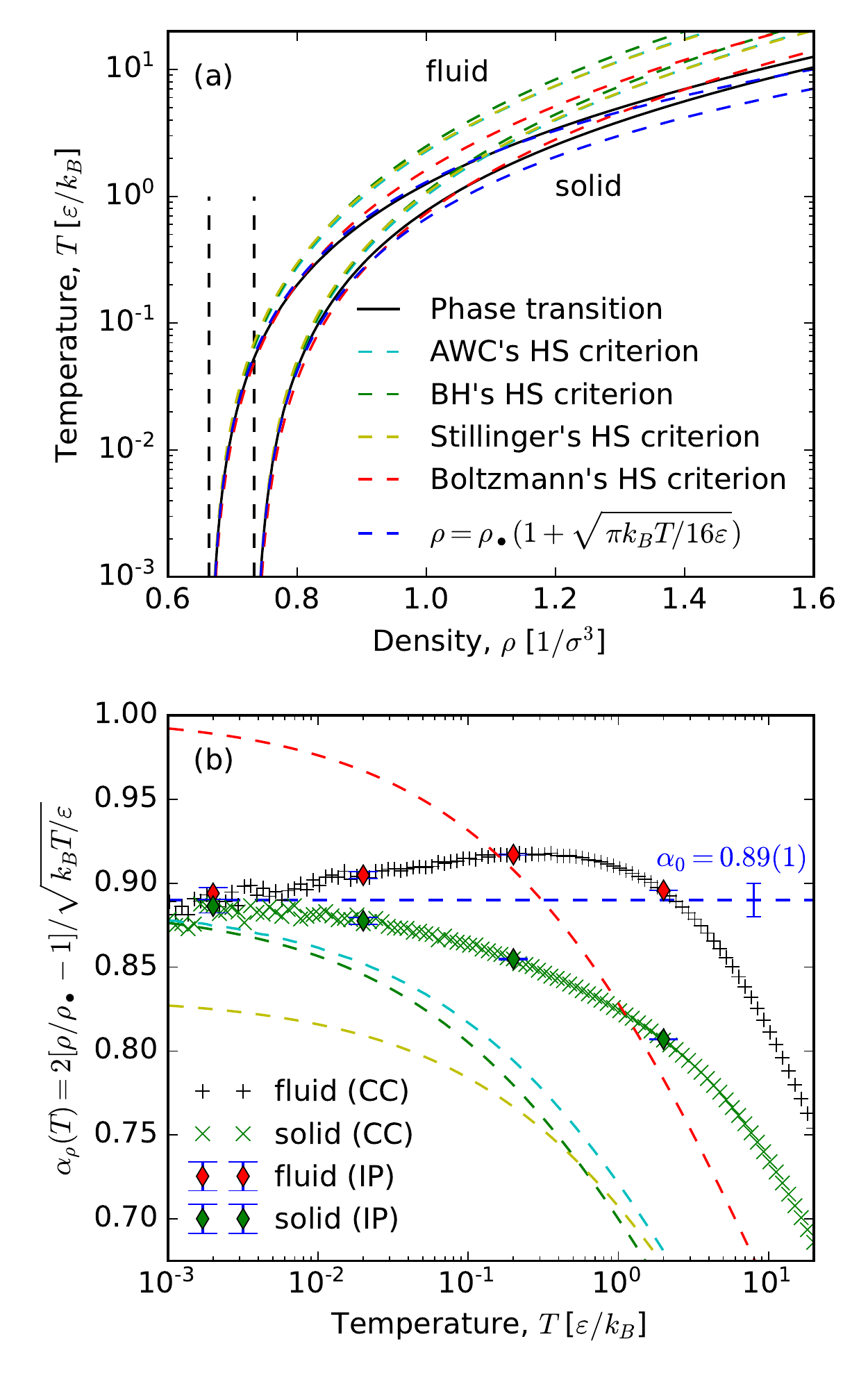}
	\caption{\label{fig:coexistence_line_rhoT} Density-temperature phase diagram. (a) The solid black lines are the coexistence densities, compare Fig.\ \ref{fig:coexistence_line_compare_rhoT}. The vertical black dashed lines mark the $T\rightarrow0$ \gls{HS} limits, i.e., the quantities $\rho_\bullet^{(l)}$ and $\rho_\bullet^{(s)}$. The turquoise, green, yellow, and red dashed curves are predictions of the \gls{HS} theories (see the text for details). The two blue dashed lines are the low-temperature fits $\rho_l=\rho_\bullet^{(l)}[1+0.445\sqrt{k_BT/\varepsilon}]$ and $\rho_s=\rho_\bullet^{(s)}[1+0.445\sqrt{k_BT/\varepsilon}]$. (b) The black $+$'s show $\alpha_\rho(T)=2({\rho}/{\rho_\bullet}-1)/\sqrt{k_BT/\varepsilon}$ where the densities $\rho$ and $\rho_\bullet$ refer to the fluid. The green $\times$'s is $\alpha_\rho(T)$ using the solid densities. The red diamonds are densities computed with the interface-pinning method. The blue error bars indicate the 95\% confidence interval. We find that the zero-temperature limit gives $\alpha_0=\lim_{T\rightarrow0}\alpha(T)=0.89(1)$. The turquoise, green, yellow, and red dashed curves are predictions of the \gls{HS} theories. The \gls{AWC} and \gls{BH} give the correct low-temperature limit within the statistical accuracy.}
\end{figure}

\section{Comparing the predictions of the different hard-sphere theories}
Having accurately located the \gls{WCA} phase transition, we can now use this to test the four \gls{HS} theories by comparing their predictions to the low-temperature \gls{WCA} melting-line data.

\subsection{Coexistence pressure and densities}
Starting with the coexistence pressure, we first need coexistence information on the \gls{HS} system. Fernandez et al. \cite{Fernandez2012} estimate that the \gls{HS} coexistence pressure is given by $p_d = 11.5727(10) k_BT/d^3$. This value is consistent with 
\begin{equation}\label{eq:HS_pressure}
	p_d = 11.5712(10)\ k_BT/d^3
\end{equation}
computed more recently by Pieprzik et al. \cite{Pieprzyk2019}; we use the latter value in this paper. In the zero-temperature limit, the \gls{HS} diameter of the \gls{WCA} interaction is $d=r_c$, which gives the coexistence pressure 
\begin{equation}
    p_\bullet = 8.1821(7)\ k_BT/\sigma^3\,.\\
\end{equation}
The bullet subscript ``$\bullet$'' refers throughout the paper to the \gls{HS} limit of the \gls{WCA} model that is approached when $T\rightarrow0$, i.e., setting $d=r_c$.

The solid black line in Fig.\ \ref{fig:HS_theory}(a) shows the coexistence pressure divided by the thermal energy, $p/k_BT$, and the black dashed line shows the $d=r_c$ prediction. The predicted pressure is too low since the effective \gls{HS} diameter is smaller than $r_c$ at finite temperature where particles are allowed to overlap. In Fig. \ref{fig:HS_theory}(a) we also consider other criteria for $d$'s (by insertions into Eq.\ (\ref{eq:HS_pressure})). At $T=0.02\varepsilon/k_B$ the $d=r_c$ criterion underestimates the coexistence pressure by 7\%, while both the \gls{AWC} and \gls{BH} criterion give only a 1\% error. Thus, the \gls{HS} theories give a significant improvement of the predicted coexistence pressure. It is hard to decide from Fig.\ \ref{fig:HS_theory} which theory is best since this depends on the temperature. We return below to the low-temperature limit that provides a definite answer. First, we turn to the \gls{HS} theories' predictions of the melting- and freezing densities.

Turning next to the freezing density, we first note that the \gls{HS} fluid freezing density has been computed recently by Moir, Lue and Bannerman to the value \cite{Moir2021}
\begin{equation}
    \rho_d^{(l)} = 0.93890(7)/d^3
\end{equation}
and the melting density of the solid to
\begin{equation}
    \rho_d^{(s)} = 1.03715(9)/d^3
\end{equation}
In the zero-temperature limit of the \gls{WCA} system ($d=r_c$) we get 
\begin{equation}
    \rho_\bullet^{(l)} = 0.66390(5)/\sigma^3
\end{equation}
and
\begin{equation}
    \rho_\bullet^{(s)} = 0.73337(6)/\sigma^3
\end{equation}
When inserting the $d$'s of the above \gls{HS} criterions we get the temperature-dependent density predictions shown in Fig.\ \ref{fig:coexistence_line_rhoT}(a) as colored dashed lines.

\subsection{Analytical treatment of the low-temperature limit}
Inspired by the functional form of Stillingers's and Boltzmann's \gls{HS} criteria (\eq{eq:d_stillinger} and \eq{eq:d_boltzmann}) we write the low-temperature limit of the effective \gls{HS} diameter as
\begin{equation}\label{eq:alpha_definition}
    d_\alpha = r_c\left(1-\frac{\alpha_0}{6}\sqrt{k_BT/\varepsilon}\right) \textrm{ for } T\rightarrow 0\,,
\end{equation}
which implies
\begin{equation}\label{eq:d3_and_alpha}
    d_\alpha^{-3} = r_c^{-3}\left(1+\frac{\alpha_0}{2}\sqrt{k_BT/\varepsilon}\right) \textrm{ for } T\rightarrow 0\,.
\end{equation}
For the Boltzmann criterion one has $\alpha_0=1$ while Stillinger's criterion gives $\alpha_0=\sqrt{\ln(2)}\simeq0.83$.

Since $d$ is the same for the \gls{AWC} and \gls{BH} criteria in the $T\rightarrow0$ limit (see Sec. \ref{sec:bh}), the $\alpha_0$'s are also identical. To evaluate $\alpha_0$ we first note that the \gls{BH} integral defining the \gls{HS} diameter (Eq. (\ref{eq:bh})) can be written 
\begin{equation}\label{eq:bh2}
    d = r_c-\int_0^{r_c}\exp(-v(r)/k_BT)dr\,.
\end{equation}
Since the \gls{WCA} pair potential is purely repulsive, it reaches its minimum at zero when $r=r_c$. Thus at low temperatures the above integral is centered near $r_c$, i.e., near $x=0$ where $x=r_c-r$. Keeping the first non-vanishing term in a Taylor expansion we get
\begin{equation}
    v(x) = \frac{1}{2}k_2x^2\textrm{ for } T\rightarrow 0
\end{equation}
with (see Ref.\ \onlinecite{Attia2021})
\begin{equation}\label{eq:k2}
k_2\equiv\left.\frac{d^2v}{dr^2}\right|_{r_c}    = 36\sqrt[3]{4}\varepsilon/\sigma^2\,.
\end{equation}
Finding $d$ from Eq.\ (\ref{eq:bh2}) involves solving a Gaussian integral in $x$. Expanding the upper limit of the integral to infinity, which is exact as $T\rightarrow 0$, we find
\begin{equation}\label{eq:ddd}
    d = r_c - \sqrt{\frac{\pi k_BT}{2k_2}}\,.
\end{equation}
By equating $d=d_\alpha$ (Eqs.\ (\ref{eq:ddd}) and (\ref{eq:alpha_definition})) we get 
\begin{equation}\label{eq:alpha_bh}
   \alpha_0 = \frac{6}{r_c}\sqrt{\frac{\pi\varepsilon}{2k_2}}
\end{equation}
or $\alpha_0=\sqrt{\pi}/2 \cong 0.886227$. The theoretical $\alpha_0$ values are summarized in Table \ref{table:alpha}.

\begin{table}
	\caption{\label{table:alpha} $\alpha_0$ values.}
	\begin{ruledtabular}
		\begin{tabular}{r|rd}
		    From simulations & & 0.89(1) \\
		    \hline
            Boltzmann & & 1 \\		    
            \gls{AWC} and \gls{BH} & $\frac{1}{2}\sqrt{\pi}=$ & 0.886227\ldots \\
            Stillinger & $\sqrt{\ln(2)}$ = & 0.832555\ldots \\
		\end{tabular}
	\end{ruledtabular}
\end{table}

To estimate $\alpha_0$ from the simulations we insert $d_\alpha^{-3}$ of Eq.\ (\ref{eq:d3_and_alpha}) into Eq.\ (\ref{eq:HS_pressure}) for the coexistence pressure, leading to 
\begin{equation}\label{eq:p_low}
    p = p_\bullet[1+\frac{\alpha_0}{2}\sqrt{k_BT/\varepsilon}]\textrm{ for } T\rightarrow 0\,.
\end{equation}
Thus, a way to determine $\alpha_0$ is to define the function (Fig.\ \ref{fig:HS_theory}(b))
\begin{equation}\label{eq:alpha}
 \alpha_p(T)=\frac{2}{\sqrt{k_BT/\varepsilon}}\left[\frac{p(T)}{p_\bullet}-1\right]
 \end{equation}
for which we note that $\alpha_0=\alpha(T)$ for $T\rightarrow0$. Similarly, we get for the densities $\rho=\rho_l$ or $\rho=\rho_s$ 
 \begin{equation}\label{eq:rho_low}
     \rho = \rho_\bullet(1+\frac{\alpha_0}{2}\sqrt{k_BT/\varepsilon})\textrm{ for } T\rightarrow 0
 \end{equation}
 and define
 \begin{equation}\label{eq:alpha_rho}
 \alpha_\rho(T)=\frac{2}{\sqrt{k_BT/\varepsilon}}\left[\frac{\rho(T)}{\rho_\bullet}-1\right]\,.
 \end{equation}
Figure \ref{fig:coexistence_line_rhoT}(a) shows the temperature dependence of the fluid and solid densities at coexistence (solid lines). These densities yield the $\alpha_\rho(T)$'s shown with black $+$'s and green $x$'s, respectively, on Fig.\ \ref{fig:coexistence_line_rhoT}(b). From the low-temperature points we estimate $\alpha_0=0.89(1)$. The colored dashed lines show the predictions of the \gls{HS} theories (the $T\rightarrow0$ limits agree with the values of Table \ref{table:alpha}). We conclude that the \gls{AWC} and \gls{BH} theories gives excellent agreement as $T\rightarrow0$. Figure \ref{fig:HS_theory}(b) shows $\alpha_p(T)$ computed using the coexistence pressure. In agreement with the results for the $\alpha_\rho(T)$'s we find that $\alpha_0=0.89(1)$ (blue dashed line). 

The success of the \gls{AWC} and \gls{BH} theories suggests writing the coexistence pressure and densities as follows (inserting $\alpha_0=\sqrt{\pi}/2$ into Eqs.\ (\ref{eq:p_low}) and (\ref{eq:rho_low}))
\begin{equation}\label{eq:p_bh_low}
    p = p_\bullet \left(1+\sqrt{\frac{\pi k_BT}{16\varepsilon}}\right)
\end{equation}
and
\begin{equation}\label{eq:rho_bh_low}
    \rho = \rho_\bullet \left(1+\sqrt{\frac{\pi k_BT}{16\varepsilon}}\right),
\end{equation}
respectively, see the blue dashed lines of Figs.\ \ref{fig:coexistence_line_compare}-\ref{fig:coexistence_line_rhoT}. Interestingly, this low-temperature approximation gives better predictions than the \emph{neat} \gls{HS} theories -- even at high temperatures (with the exception of Boltzmann's criterion near $T\simeq0.5\varepsilon/k_B$). We do not have an explanation for this.

Equations (\ref{eq:p_bh_low}) and (\ref{eq:rho_bh_low}) summarize an important result of this paper, providing an analytical \gls{HS} approximation for the low-temperature freezing of the \gls{WCA} fluid. This can be generalized to any other purely repulsive pair-potential that is truncated smoothly at $r=r_c$ by the following steps:
\begin{enumerate}
    \item Compute $k_2$ using Eq.\ (\ref{eq:k2}) and
    \item derive $\alpha_0$ within the \gls{BH} theory by inserting $k_2$ into Eq.\ (\ref{eq:alpha_bh}).
    \item Low-temperature predictions for coexistence pressure and densities are then provided by inserting $\alpha_0$ into Eqs.\ (\ref{eq:p_low}) and (\ref{eq:rho_low}), respectively.
\end{enumerate}

\begin{figure}
	\includegraphics[width=1.0\columnwidth]{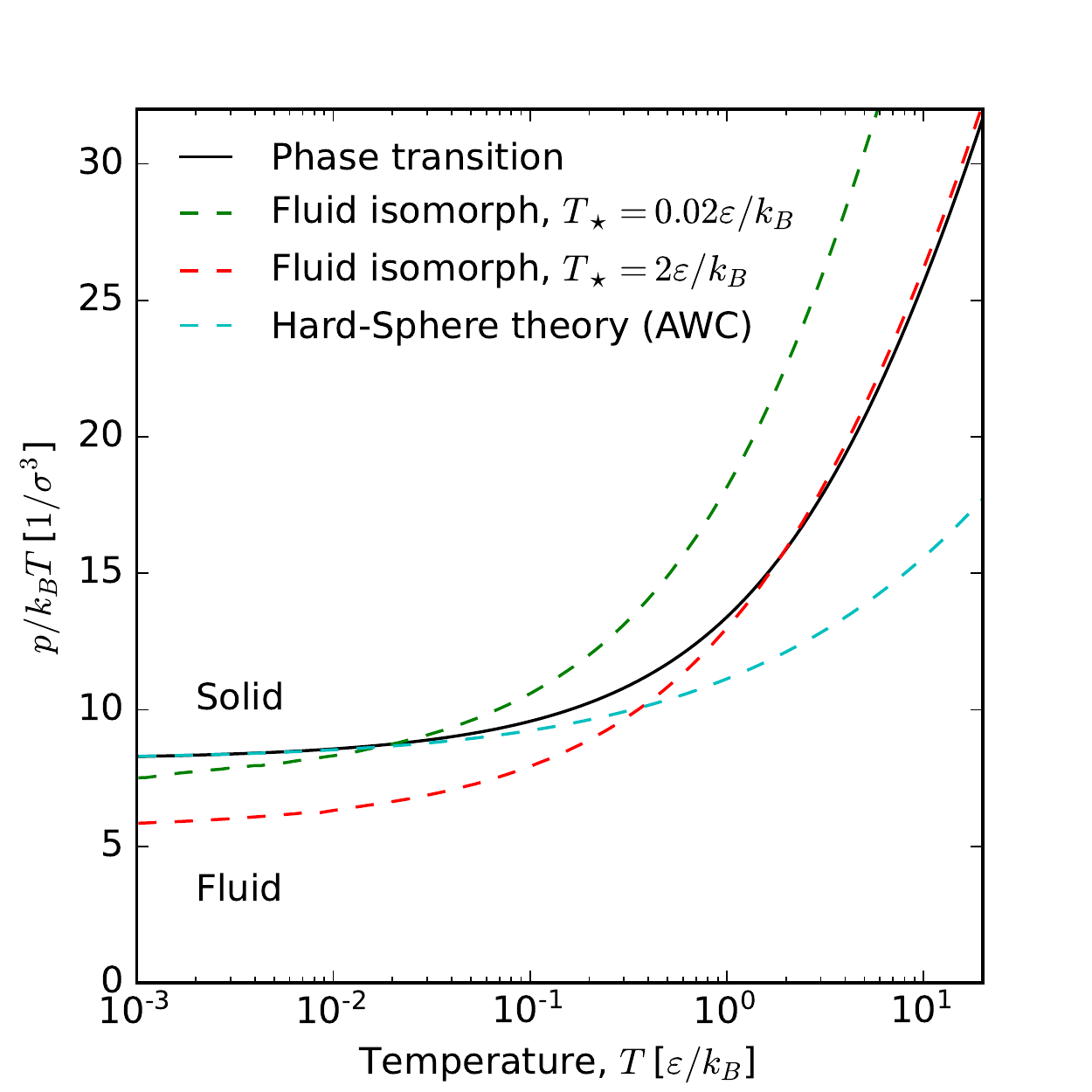}
	\caption{\label{fig:isomorph} The solid black line shows the reduced coexistence pressure $p/k_BT$ as a function of the temperature. The red and green dashed curves are isomorphs of the fluid, i.a., lines along which the excess entropy, $S_{ex}$, is constant. The isomorphs touch the phase-transition line at $T_\star=0.02\varepsilon/k_B$ and $T_\star=2\varepsilon/k_B$, respectively. The turquoise dashed line is the prediction of the \gls{AWC} theory.}
\end{figure}

\begin{figure*}
	\includegraphics[width=2.0\columnwidth]{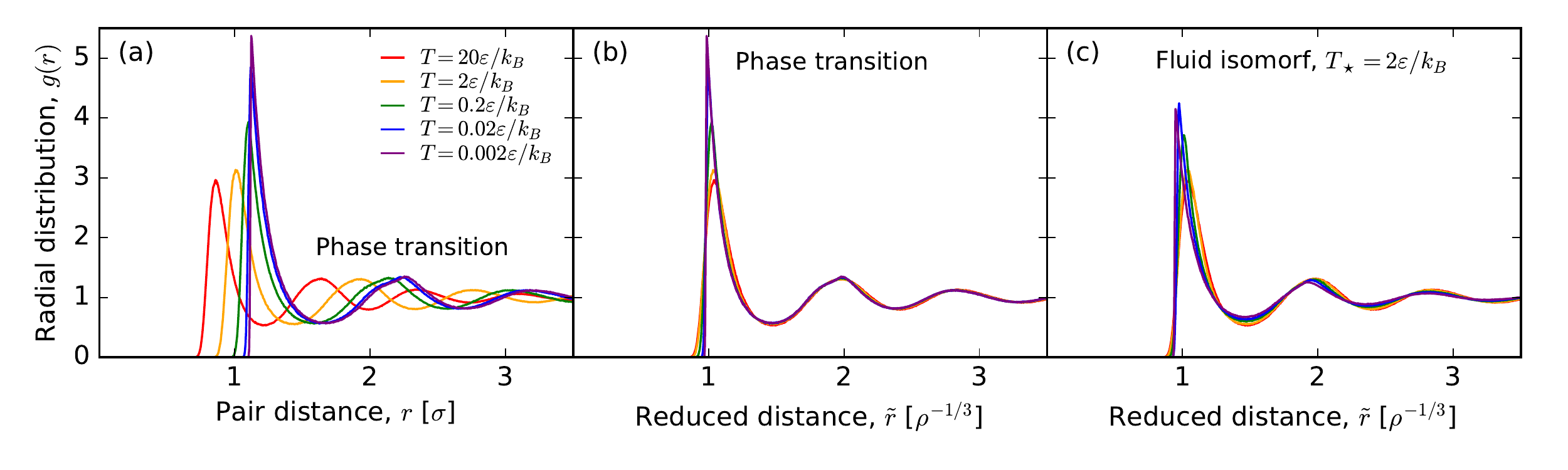}
	\caption{\label{fig:rdf} (a) The radial distribution $g(r)$ for the fluid at coexistence. (b) The radial distribution as a function of the reduced distance $\tilde r = r\sqrt[3]{\rho}$ for the fluid at coexistence. (c) The radial distribution as a function of the reduced distance $\tilde r = r\sqrt[3]{\rho}$ for a fluid isomorph that touch the coexistence line at $T_\star=2\varepsilon/k_B$. }
\end{figure*}

\begin{figure}
	\includegraphics[width=1.0\columnwidth]{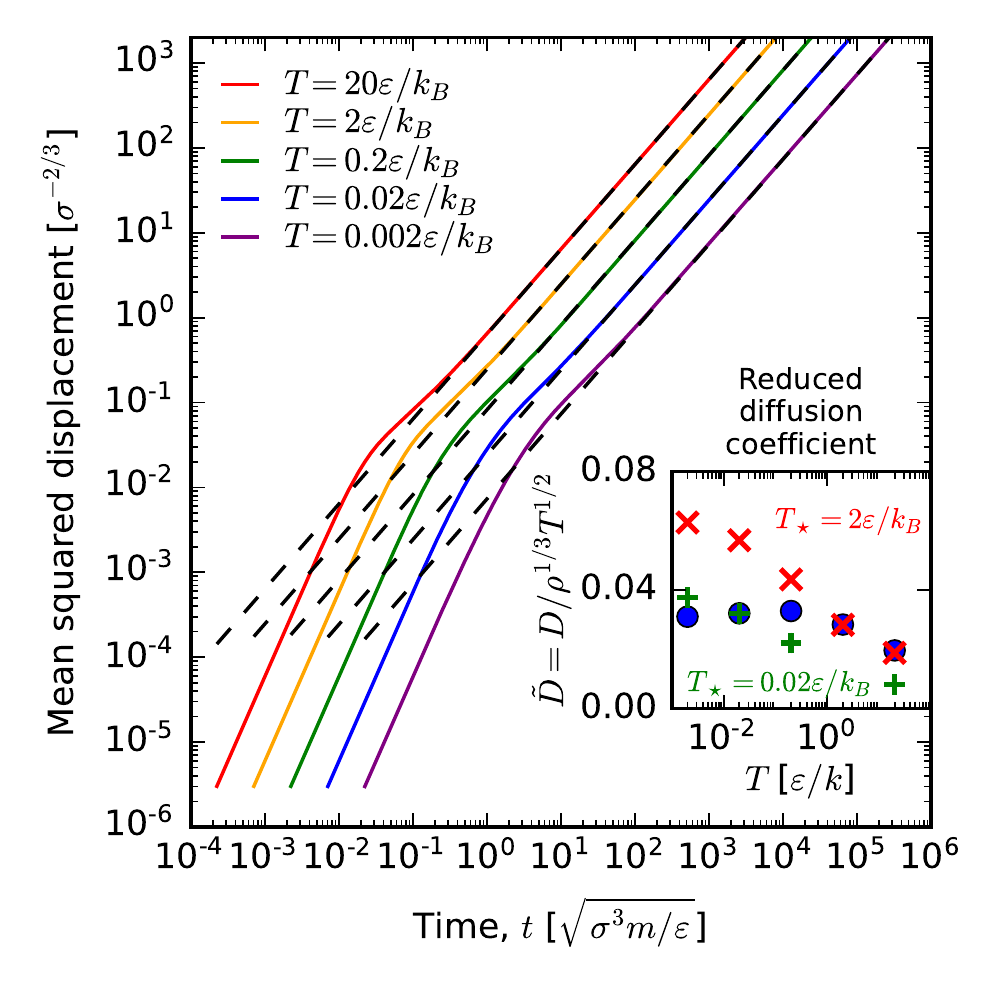}
	\caption{\label{fig:msd} The solid lines show the mean-square displacement $\langle |{\bf r}(t)-{\bf r}(0)|^2\rangle$ for selected state-points along the coexistence line (see Table \ref{table:phase_transition}). The dashed lines are long-time fits to $\langle |{\bf r}(t)-{\bf r}(0)|^2\rangle=6Dt$ where $D$ is the diffusion coefficient. The dots in the inset show the reduced diffusion coefficient $\tilde D=D\rho^{1/3}T^{1/2}$. The red $x$'s and green $+$'s is the reduced diffusion coefficient for state points along the isomorphs with $T_\star=2\varepsilon/k:B$ and $T_\star=0.02\varepsilon/k_B$, respectively.}
\end{figure}

\section{Outlook}

We have shown that \gls{HS} theories give excellent predictions of the \gls{WCA} melting line at low temperatures, in particular for the \gls{AWC} and \gls{BH} approximations. At higher temperatures the \gls{HS} theories are less accurate. This is not surprising because the \gls{WCA} model only resembles a \gls{HS} system at low temperatures. How to predict the \gls{WCA} melting-line pressures and coexistence densities at high temperatures? One possibility is to generalize the low-temperature \gls{HS} approximation by considering the lines of constant excess entropy $S_\textrm{ex}$ (this is the entropy in excess of the ideal gas entropy at the same density and temperature, a negative quantity that in some textbooks \cite{Prigogine1954} is referred to as the residual entropy). For the \gls{HS} system these lines are determined entirely by the density, i.e., they are vertical in the density-temperature phase diagram. In Ref.\ \onlinecite{Attia2021} it has been shown that the \gls{WCA} system's structure and dynamics are near-invariant along the lines of constant excess entropy, which are referred to as isomorphs \cite{Gnan2009,Schroeder2014}. An isomorph can be computed by numerical integration in the $\ln T$-$\ln \rho$ plane (e.g., using the fourth-order Runge-Kutta method (RK4) \cite{Attia2021}) for which the required slope is $f=1/\gamma$ where \cite{Gnan2009,Dyre2014,Dyre2018}
\begin{equation}\label{eq:gamma}
	\gamma\equiv\left(\frac{\partial\ln T}{\partial\ln \rho}\right)_{S_\textrm{ex}}\,.
\end{equation}
The ``density-scaling exponent'' $\gamma$ may be computed from virial- and potential-energy fluctuations in the $NVT$ ensemble via the general statistical-mechanical identity  $\gamma=\langle \Delta W\Delta U\rangle/\langle (\Delta U)^2\rangle$ \cite{Gnan2009}. Figure \ref{fig:isomorph} shows the reduced pressure $p/k_BT$ of two fluid isomorphs that overlap with the coexistence line at $T_\star=0.02\varepsilon/k_B$ and $T_\star=2\varepsilon/k_B$, respectively (dashed green and red lines). For comparison, the turquoise dashed line shows the prediction of the reduced coexistence pressure of the \gls{AWC} theory. For the entire temperature span the isomorphs gives predictions with an overall accuracy comparable to that of the best \gls{HS} approximation (\gls{AWC}). 

Figures \ref{fig:rdf} and \ref{fig:msd} show the structure and dynamics along the melting line and the fluid isomorph in reduced units \cite{Gnan2009}. Interestingly, the physics is more invariant along the coexistence lines than along the isomorph. This is in contrast to previous findings for the LJ system, where the opposite applies \cite{Pedersen2016}. We note, however, that isomorphs only follow the coexistence lines to a first approximation. For the LJ system, accurate predictions for the thermodynamics of freezing and melting can be arrived at within the isomorph-theoretical perturbation framework proposed in Ref.\ \onlinecite{Pedersen2016} -- we hope to apply the same method to the \gls{WCA} system in the near future.

\section*{Acknowledgment}
This work was supported by the VILLUM Foundation’s
Matter grant (No. 16515).

\section*{Author Declarations}
\subsection*{Conflict of Interest}
The authors have no conflicts to disclose.

\section*{Data availability}
The data that support the findings of this study are openly available in Zenodo at \url{http://doi.org/10.5281/zenodo.6505218}, reference number 6505218.

\bibliography{references}
\end{document}